# Pythia: Scheduling of Concurrent Network packet Processing Applications on Heterogeneous Devices [EXTENDED VERSION]

Giannis Giakoumakis
*ICS-FORTH*
Heraklion, Crete, Greece
ggiakoum@ics.forth.gr

Eva Papadogiannaki
*ICS-FORTH*
Heraklion, Crete, Greece
epapado@ics.forth.gr

Giorgos Vasiliadis
*ICS-FORTH*
Heraklion, Crete, Greece
gvasil@ics.forth.gr

Sotiris Ioannidis
*ICS-FORTH*
Heraklion, Crete, Greece
sotiris@ics.forth.gr

*Abstract*—Modern commodity computing systems are composed by a number of different heterogeneous processing units, each of which has its own unique performance and energy characteristics. However, the majority of current network packet processing frameworks targets only a specific processing unit (either the CPU or accelerator), leaving the remaining computational resources under-utilized or even idle.

In this paper, we propose an adaptive scheduling approach for network packet processing applications, that supports any heterogeneous and asymmetric architectures that can be found in a commodity high-end hardware setup. Our scheduler not only distributes the workloads to the appropriate devices in the system to achieve the desired performance results, but also enables the multiplexing of diverse network packet processing applications that execute concurrently, eliminating the interference effects introduced at runtime. The evaluation results show that our scheduler is able to tackle interferences in the shared hardware resources as well to respond quickly to dynamic fluctuations (e.g., application overloads, traffic bursts, infrastructural changes, etc.) that may occur at real time.

## I. Introduction

The advent of high-end commodity heterogeneous systems (i.e., systems that utilize multiple processing units, typically CPUs along with different types of GPUs) has motivated the networking community to exploit alternative architectures. In fact, many recent approaches utilize them appropriately to build high-performance and parallel network packet processing systems [17], [32], [40], as well as power efficient systems [29]. Unfortunately, the majority of these approaches often target a single computational *device* [1], such as a multi-core main processor or a powerful high-end GPU, excluding the remaining devices, leaving them completely idle. Developing a network processing application framework that can utilize *each* and *every* available device effectively, efficiently and consistently, between a wide range of diverse workloads running concurrently, is highly challenging.

Heterogeneous systems that consist of multiple devices, typically provide system designers with different optimization opportunities that could eventually introduce inherent constraints and trade-offs between energy consumption and other performance characteristics — in our case, forwarding throughput and latency. The challenge to fully utilize a heterogeneous system, is to map the requested computations to the processing devices that interfere the less, and do it in the most automated way possible. Previous works focused on developing load-balancing frameworks that automatically partition the workload across the devices [8], [21].These approaches either assume that all devices can provide equal performance [21] or perform a series of small processing trials to determine their relative performance [8]. The major disadvantage of these approaches is that they have been designed for solo applications, i.e., only a single application is executing each time. However, this is not the case for networking middleboxes, in which the complexity constantly increases, requiring more and more networking applications to execute at the same time. In addition, the majority of these approaches take as input a constant stream of data, a limiting factor that force them to adapt poorly when the input stream rates vary. This makes them hard to apply to network environments, where the traffic variability [7], [26] and overloads [9] can significantly affect the utilization and performance of network applications.

In this paper, we propose Pythia [14], which is a scheduling approach for network packet processing applications that can be executed concurrently in a highly heterogeneous commodity base system. More specifically, our proposed scheduler is designed to explicitly focus on the heterogeneity that is introduced in (i) the underlying hardware architectures, (ii) the applications and (iii) the input network traffic rate. The scheduler can dynamically respond to dynamic performance fluctuations that can occur at any time during the runtime, such as traffic bursts, overloads and system changes. The contributions of this work are the following:

- We measure the performance achieved as well as the power consumption of several typical software network packet processing applications that can be executed concurrently on commodity heterogeneous systems. We show how the combination of different set of devices (i.e. CPUs, integrated GPUs and discrete GPUs) can affect the performance and we discuss the problems that

---
[1]Hereafter, we use the term "device" to refer to computational devices, such as a CPU and a GPU, unless explicitly stated otherwise.

- can arise by their concurrent utilization, specifically for network packet processing applications.
- We show that the performance results have wide variations when executing diverse types of network processing applications. For several cases, a specific device can be the best fit for one application type, while at the same time, it can be the worst choice for another.
- We present an energy profiling tool that reports live power consumption measurements for any device in a system setup, reading the corresponding hardware performance registers.
- Motivated by the current gap in the state-of-the-art, we present a scheduling approach that, given a set of network packet processing applications, can effectively and efficiently utilize the most appropriate device or group of devices, based on the current system and network conditions, using a predefined policy that specifies the performance goal. The scheduler is able to dynamically respond to system and performance fluctuations, and provide consistently good performance for concurrently running applications.

## II. BACKGROUND

Traditional commodity hardware setups offer a three-level heterogeneity: (i) the x86 CPU architecture, (ii) the integrated GPU that is packed on the same processor chip, and (iii) a discrete high-end GPU. All these different hardware architectures offer unique performance rankings and diverse energy characteristics. CPU cores perform overall better under branch-intensive workloads, while discrete GPUs perform efficiently in data-parallel tasks. An integrated GPU offers low power consumption with a fair computational rate and relatively low latency. Typically, a discrete GPU is connected with the main processor through the PCIe bus and they do not share physical address space. An integrated GPU, on the other hand, shares the LLC cache and the memory controller with the CPU.

In Figure 1(a), we illustrate the packet processing scheme that has been used by approaches that utilize a discrete GPU [19], [37], [38], [40]. The majority of these approaches perform the following steps [2]: (i) the DMA transaction between the NIC and the main memory, (ii) the transfer of the packets to the I/O bus that corresponds to the discrete

[2]We assume that a packet batch is already in the NICs internal queue.

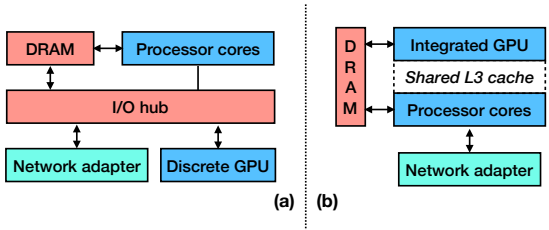

Fig. 1. Packet processing on different GPU architectures.

GPU (iii) the DMA transaction to the memory space of the discrete GPU, (iv) the actual processing GPU kernel itself and (v) the transfer of the results back to the host memory. Due to the PCIe interconnect inability to quickly handle small data transfers, all data transfers must operate on large batches. An equivalent architecture is illustrated on the right side of Figure 1, using an integrated GPU that is packed on the same die with the CPU. The advantage of this approach is that the integrated GPU and CPU share the same physical memory address space, which allows in-place data processing, resulting to lower execution latency. As we observe through our experiments, this hardware architecture has lower power consumption (e.g. even the absence of the I/O Controller Hub alone saves energy), when compared to the discrete GPU setup of Figure 1(a). Compared to high-end discrete GPUs, integrated GPUs (i) are typically built using low-power, 3D transistor manufacturing process, (ii) have a simple internal architecture and no dedicated main memory to add extra power consumption, and (iii) they match the computational requirements of applications, in which the main bottleneck is the I/O interface and thus, a discrete GPU would be underutilized.

## III. SYSTEM SETUP

In this section we describe the hardware platform that we use in our measurements, as well as our power consumption profiling tool. In addition, we discuss the network packet processing applications that we use in this work.

### A. Hardware Setup

Our hardware setup contains one Intel Core i7-8700K Coffee Lake processor, packed with a UHD Graphics 630 graphics card, and one high-end NVIDIA GeForce GTX 1080 Ti graphics card. The processor contains six CPU cores (3.7GHz), with hyper-threading support that results in twelve hardware threads. Overall, our machine contains three different, heterogeneous, commodity hardware devices: one CPU, one integrated GPU and one discrete GPU. The system is equipped with 32GB of dual-channel DDR4-2666 DRAM with 41.6 GB/s throughput. The L3 cache is 12MiB, and along with the memory controller, it is shared across the CPU cores and the integrated GPU. Each CPU core is equipped with 192KiB of L1 cache and 1.5MiB of L2 cache. The GTX 1080 Ti has 3584 cores and 11GB of GDDR5X memory. It is rated at 11.34 TFlops and its Thermal Design Power (TDP) is 250 Watt. The UHD Graphics 630 has 24 execution units and its maximum performance is estimated at 423.2 GFlops (1150Mhz). The TDP for the whole processor is 95 Watt. The motherboard that is included in our setup is a Gigabyte Z370 AORUS Ultra Gaming. Our three-way heterogeneous hardware setup presents an interesting trade-off: even though the integrated GPU has fewer and less powerful resources (e.g. execution units, hardware threads), when compared to a high-end discrete graphics card, the integrated GPU has satisfying performance, especially since it consumes much lower power due to the fact that it is directly connected to the processor and system's main memory via a fast on-chip ring bus.

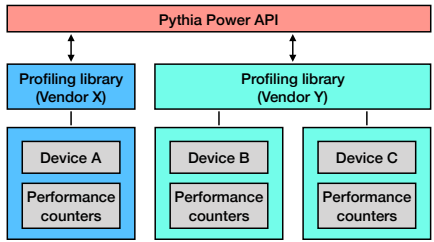

Fig. 2. Pythia Power API.

### B. Energy & Power Profiling

The divergence of the types and the vendors of the hardware devices introduces a problem when measuring hardware statistics, such as energy and power consumption. To overcome this obstacle, we implement a standalone tool that reads hardware performance registers of each device and periodically reports the consumed energy for a given time interval. The profiling tool is dependent on the main hardware profiling library of each vendor, such as NVIDIA's NVML and a customized version of Intel's PCM. The implementation of the profiling tool is straight forward. We provide an API with wrapper functions that internally decide which vendor library must be used, based on the type of the device that is being profiled. Our proposed tool is completely detached from the OpenCL framework, and thus, it can be used by any application. Figure 2 presents an overview of the Pythia Power API that we developed.

### C. Kernel Multiplexing

In modern commodity hardware it is possible to achieve parallel execution of more than one kernels by using the OpenCL Framework. However the way of splitting hardware resources to two or more kernels is usually dependent on the type of the device. We put different GPU devices under test and found out that by just creating a discrete command queue for each different kernel, it is possible to interleave their execution. However, we have no information or control over the execution of each kernel, such as how many compute units each kernel will be assigned and when context switches will occur. The hardware of the device takes care of such details. On the other hand, there is an OpenCL extension, allowing us to split a CPU device in many different ways. In other words, we have full control over how many and which exactly compute units will be assigned. We can also specify how many (if any) compute units will remain unused by the OpenCL framework, in order to be used by the host program.

### D. Applications

Taking advantage of the uniform execution that the OpenCL framework offers [1], we implemented three typical packet processing applications, which are commonly deployed in network appliances and involve both processing and memory-intensive behaviours.

*a) Deep Packet Inspection:* Deep packet inspection (DPI) is a common operation in network traffic processing applications. It is used in traffic monitoring and classification tools, as well as in network intrusion detection systems, antiviruses etc. In our implementation, we use the Aho-Corasick algorithm [5] that offers simultaneous multi-pattern searching. We use 10,000 patterns of fixed strings that come from the latest signatures of Snort [3] distribution. The patterns are compiled into a DFA automaton. For our performance measurements below, we generate network traffic using the netmap tool, namely pkt-gen, and we inject it with content that results to around 30% match reporting.

*b) Packet Hashing:* Packet hashing is used in redundancy elimination and in-network caching systems [4], [6]. Redundancy elimination systems traditionally maintain a "packet store" and a "fingerprint table". The packet store gets updated right after a packet reception, and the fingerprint table is being checked to determine whether the packet includes an important fraction of content cached in the packet store. In that case, an encoded version gets transmitted, eliminating this recently observed content. Specifically, we have implemented the MD5 algorithm. MD5 presents low probability of collisions and it is mainly used for checking data integrity [2] or deduplication [23].

*c) Encryption:* The third application that we develop is the AES-CBC encryption mechanism using 128-bit keys per connection. Encryption is used by protocols and services, such as SSL, VPN and IPsec, for securing connections by authenticating and encrypting the packets inside a communication channel. By employing end-to-end encryption, we can protect data flows between pairs of hosts, security gateways, etc. Due to its nature, this encryption technique is a representative form of computational-intensive packet processing.

## IV. ARCHITECTURE

In this section we describe two different architectural models for our system. The main difference between the two models is the way each one handles the incoming network traffic and distributes it to the computational devices for further processing.

### A. Master/Worker Architecture

In this approach our system creates two sets of threads, with each set having a separate role. The first set of threads, the master threads, periodically poll the network interfaces, capture network packets from the driver's receiving queues and fill input data buffers. Each of those buffers is shared between a master thread and a worker thread, which is responsible for polling the corresponding data buffer and decide when there are enough packets in order to transfer it to a target device and spawn the device execution. In the meanwhile, the masters should keep filling buffers with new packets and repeat the procedure upon target device execution completion. In order to be able to achieve such pipeline, we create our buffers as follows: there is (i) an input buffer, (ii) a swap buffer and (iii) an output buffer. A master thread is responsible for filling the corresponding input buffers, then when a worker decides to spawn device execution, it swaps the contents of the input buffer to the swap buffer, it transfers the processed data from the target device back to the host program using the output buffer and finally transfers the swap buffer to the target device and spawn again its execution on the new input data. Using such a structure for the buffers enables the master threads to keep capturing packets and filling buffers even when devices are busy with the processing procedure. In our setup we use the netmap module to reduce the cost of packet copying from the network interfaces to the input buffers. Netmap only supports the use of up to 4 interfaces, so just one master thread is enough. In our configuration, the master thread is pinned in a CPU core and the utilization due to interrupts in order to capture network traffic is between 40% (if only one network interface is active) and 60% (if all four interfaces are transmitting). The rest of the CPU cores are either workers or perform actual processing, if CPU is also a target device. If we use a different module to capture network traffic and add some extra network interfaces this model introduces serious scalability problems as a single master thread is no longer capable of capturing the traffic from all network interfaces. More than one masters will be spawned and the processing capacity for the workers will decrease. The architecture is displayed in Figure 3(a).

*B. Lock-free Architecture*

A major disadvantage of the master/worker architecture, is the need of synchronization between the master thread and each worker in pairs. In addition, the master thread is also in charge of handling the input traffic from the network interfaces. In our hardware setup, the master thread occupies 60% of its CPU core. However, adding even more network interfaces, would require the existence of multiple master threads that would lead to a demand of more complex synchronization between each other. Of course, scalability is of major importance for such systems. Thus, we implement a second model, with worker threads that do not rely on masters in order to consume the incoming network packets. In this new approach each worker is responsible for three main procedures: (i) capture network traffic from a discrete set of network interfaces, (ii) spawn the execution on the target device and (iii) collect statistics of the most recent execution cycle of the target device. Thus, instead of relying on a master thread to collect all the received network traffic and distribute it accordingly, we bind a single worker with a separate network interface, as shown in Figure 3(b). Then, the number of the workers is equal to the number of the available network interfaces. This architecture splits the incoming traffic across the main processor cores, and each one is responsible to perform all the computations for every packet it receives in a lock-free manner. Compared to the master/worker architecture, the lock-free model alleviates the overhead caused by the synchronization burden, which is essential to ensure the normal execution of the workers, as well as the correctness of the results.

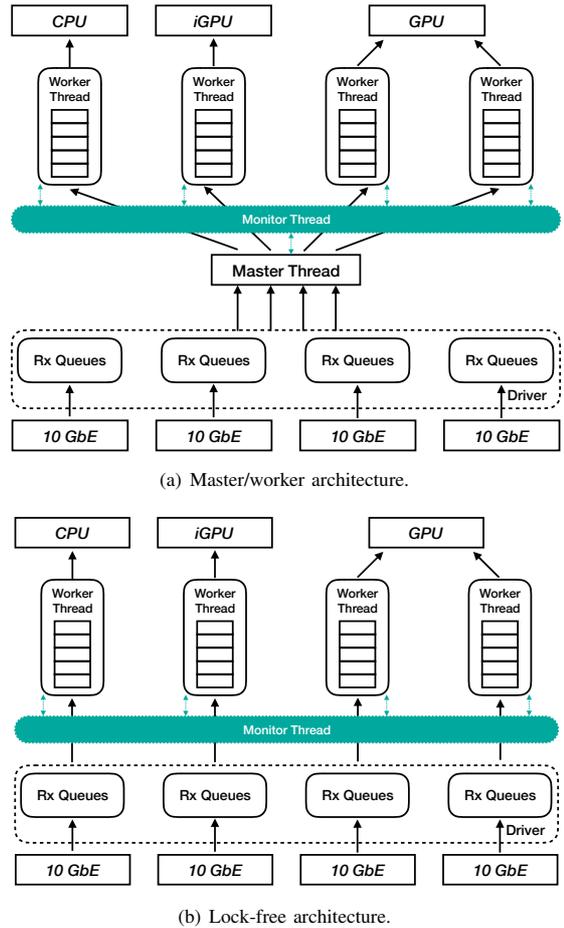

(a) Master/worker architecture.

(b) Lock-free architecture.

Fig. 3. Master/worker versus Lock-free architecture for capturing network packets and distributing them to the available computational devices.

## V. IMPLEMENTATION

To uniformly execute all the implemented packet processing applications (§ III-D) across every device in our testbed machine, we take advantage of the OpenCL framework. Our system runs Linux 4.19.34-1-lts. We use the Intel OpenCL 2.1 SDK for the Intel devices (i.e. the i7-8700K CPU and the UHD Graphics 630 GPU) and the OpenCL from the NVIDIA CUDA Toolkit 10.1 for the GTX 1080 Ti GPU.

Each of the three applications is implemented as a unique kernel. In OpenCL, an instance of a kernel is called *work-item* and a set of multiple work-items is called *work-group*. We follow a packet-per-thread approach (such as other relevant works [12], [17], [40]). This means that each work-item reads at least one packet from the memory and then performs the packet processing. Different work-groups can run concurrently on different hardware cores. The management of work-groups can present an interesting performance trade-off; a large num-ber of work-groups offers more flexibility in scheduling, something that increases the switching overhead, though. Typically, GPUs contain a significantly faster thread scheduler, thus it is recommended to spawn a large number of work-groups, since it hides the latency that is introduced by heavy memory transfers through the PCIe bus. While a group of threads waits for data consumption, another group can be scheduled for execution. On the other hand, CPUs perform more efficiently, when the number of work-groups is close to the number of the available cores. Discrete, high-end GPUs have a dedicated memory space, physically independent from the main memory. To execute a task on the GPU, we must explicitly transfer the data between the host (i.e. CPU) and the device (i.e. GPU). Data transfers are performed via DMA, so the main memory region should be page-locked to prevent any page swapping during the time that transfers take place. In OpenCL, a data buffer, which is required for the execution of a computing kernel, has to be created and associated with a specific *context*. Different contexts cannot share data directly. Thus, we have to explicitly copy the received network packets to a separate, page-locked buffer that has been allocated from the context of the discrete GPU and can be moved towards its memory space, safely via PCIe. Data transfers (host-device-host) and GPU execution are performed asynchronously, permitting a pipeline of computation and communication, something that significantly improves parallelism. Through our performance measurements, we notice that different applications have different data transfer requirements. For instance, applications like DPI and MD5 do not change the headers or the payloads of the packets, so there is no need to transfer them back to the host after the execution. On the other hand, packets need to be transferred back to the main memory, when processed by the AES application, since it changes their contents. Still, when the processing is performed on the main processor or an integrated GPU, expensive data transfers are not required, since both devices have direct access to the host memory. To avoid redundant copies, we explicitly map the corresponding memory buffers between the CPU and the integrated GPU, using the OpenCL's `clEnqueueMapBuffer()` function.

Memory accesses can be critical to the overall performance sustained by our applications. For instance, the memory loads in GPUs are more effective when data are in a column-major order (the so-called memory coalescing [30]). CPUs, on the other hand, require row-major order to benefit from cache locality within each thread. As these two patterns are contradictory, we transpose the whole packets to column-major order to benefit from memory coalescing, only when reside within the GPU memory. The overall costs, however, pay off only when accessing the memory with small vector types (i.e. `char4`). When using the `int4` type though, the overhead is not amortized by the resulting memory coalescing gains in none of our representative applications (similar to [31]). Besides GPU, the CPU also offers support for single-instruction-multiple-data (SIMD) units when using the `int4` type, since the vectorized code is translated to SIMD instructions [33]. Therefore, we access the packets using `int4` vector types in a row-major order, for both hardware architectures (i.e. the CPU and the GPU). Furthermore, OpenCL offers the so-called *local memory*, which is a memory region that is shared by every work-item inside a work-group. This local memory is implemented as an on-chip memory on GPUs, which is much faster than the off-chip global memory. Hence, GPUs take advantage of this local memory to improve performance. On the other hand, CPUs do not have any similar physical memory region to be used as local memory. As a result, all memory objects located inside the local memory are mapped to sections of the global memory, a procedure that causes performance overheads. To solve this, we explicitly put data to local memory only when the computations are meant to be performed on the discrete GPU.

### A. Batch Processing

Network packets are placed into batches exactly in the same order they are received through the network interfaces. When multiple devices are used simultaneously, packets can be reordered. To prevent packet reordering we synchronize devices using a barrier, enforcing all involved devices to execute in a lockstep fashion. This approach, though, would reduce the overall performance of a system, as fast devices would be forced to wait for the slow ones. This can be a major drawback in setups where the devices have high computational capacity differences. To bypass this problem, we firstly classify incoming packets by building the typical 5-tuple flows before creating the batches, and then we ensure that the packets that belong to the same flow will never be placed in batches that will be processed simultaneously by different devices. Batches get delivered to the corresponding devices, by the CPU core that manages the traffic coming from the corresponding network interface. Each device has a different queue, allocated inside the device's context that newly arrived batches of packets are inserted.

### B. Performance Measurements

We now present the performance achieved by different configurations, created by combinations of device(s) and application(s) executed in our hardware setup. We use the netmap framework [32] to generate and transmit network packets to our machine[3]. Due to space constraints, we present only the performance achieved using a small fraction of the possible configurations. However, we carefully select these configurations to better show the diverse performance characteristics of each device and application[4]. To get power measurements, we use our power instrumentation tool and we report the aggregated power consumption of the distinct devices that are used for the processing. To accurately measure the power consumption of each device processing the corresponding batch, we measure the power spent for every unique device that is required for the execution. For example, when the discrete GPU is used for packet processing, the CPU is not idle, since it has to collect the necessary packets, transfer them to the device dedicated memory, spawn the kernel for execution, and then transfer the results back to the host memory. On the other hand, when we use only the CPU (or the integrated GPU) for processing, we power-off the discrete GPU, since it will be unused.

---

[3]Through the performance measurements and the evaluation of our scheduler, we use the netmap framework to exchange network traffic between two interconnected end-hosts through a 40-Gbps NIC (4 interfaces of 10 Gbps each). However, the overall throughput achieved is not higher that 30 Gbps; the reason is that the motherboard used in our hardware setup does not support two x16 PCIe slots concurrently, hence both our NIC and discrete GPU (GTX 1080 Ti) run at a reduced I/O bandwidth (x8).

[4]In the presented performance measurements, we do not use batch sizes larger than 16K packets (e.g. 32K, 64K, etc.), since the maximum throughput observed can be achieved using 16K batch size.

Tables I, II and III summarize a proportion of the possible configurations and the performance achieved. Table I presents both the individual and the aggregated performance achieved by the DPI application, when executed either standalone or by sharing the device with 1 or 3 co-workers. The same benchmark executions are repeated for all the available devices of the system, i.e. i7-8700K, UHD graphics card and the GTX 1080 Ti. Similarly, Tables II and III display the individual and the aggregated performance achieved by the AES and MD5 applications respectively, when executed standalone or alongside some co-workers on the same devices as in Table I. We note that the current implementation of our scheduler supports the concurrent execution of every network packet processing application combined; for the purposes of simplicity though, we present only the combination of two different applications in each device at a time. When we utilize multiple devices or when we execute more than one concurrent applications, the batch of packets needs to be further split into sub-batches of different size that will be properly offloaded. We benchmarked all possible combinations of packet batches, devices and applications. Due to space constraints though, we plot only the average performance achieved for each case. In the case of the CPU processor, we include the results using all six cores (twelve logical threads) in parallel. There are other works that use a Xeon processor, configured with two NUMA nodes. Instead, we use a single-node i7 CPU to take advantage of the integrated GPU that is packed in the same processor die. However, we strongly believe that our system will reach a doubled performance with a dual-node configuration.

Via our experiments, we observe that the throughput is always improved while increasing the batch size. However, different applications require a diverse batch size to reach their maximum throughput. Processing intensive applications (i.e. AES) benefit more from large batch sizes, while memory intensive applications (such as the DPI application) can reach the peak throughput using smaller batch sizes. For example, for the DPI in Figure 4 we see that a working set larger than 4096 packets results in lower overall throughput for the CPU. Increasing the batch size after the maximum throughput has been reached, results to linear increases in latency. Furthermore, we can also notice that the sustained throughput is not consistent across diverse devices. For instance, an integrated GPU seems to be a reasonable choice when performing MD5 and DPI on large batches of packets, compared to AES, where the integrated GPU results to low throughput. A CPU is the best option for latency-aware environments, using a small batch size. Apparently, there is not clear ranking between the devices, not even a clear winner. As a matter of fact, there are devices that can be the best fit for some applications, while at the same time, the worst option for another.

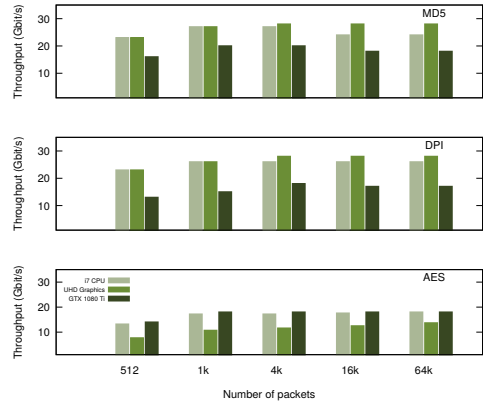

Fig. 4. Throughput of the different devices that comprise our base system executing the following applications for 1500-bytes network packets: (a) MD5 hashing, (b) Deep Packet Inspection, and (c) AES-CBC encryption

When executing concurrently more than one network packet processing applications in one device, we reach the challenge of unknown interference effects. These effects include but are not limited to: contention for hardware resources (e.g., shared caches, I/O interconnects, etc.), software resources, and false sharing of cache blocks. A major difference between single versus multiple concurrent application execution on the discrete GPU reveals an interesting finding. Batch size should be chosen wisely based on the number of applications concurrently executed as a large batch size (16K) has negative effects in cases where more than one applications are being offloaded to the GTX 1080 Ti. The reason behind this is the fact that both the discrete graphics card and the NIC, share the same I/O interconnect (i.e., the PCI bus), as explained below. Another interesting fact has to do with multiple instances of AES on the same device. When this is the case, the aggregated performance is lower compared to the aggregated performance of every other kernel combination on a given device, as shown in Table II. GTX 1080 Ti is an exception as it is not affected by the compute-intensive nature of AES and is able to sustain top performance even when four instances of AES are executed at the same time. On that note, despite that the integrated graphics card performs tolerably well on single AES execution when combined with a large batch size (Figure 4), it is the least suitable device when the desired scenario requires the concurrent execution of an AES instance alongside an instance of any other application, as shown in the bottom part of Table II. Moreover, when DPI is coupled with MD5 (Tables I and III), the GTX 1080 seems to be the worst fit; not only giving the worst performance but it does so by being the most energy-hungry device at the same time. The CPU and the integrated GPU both achieve similar performance results, but in the case of the UHD graphics card, top performance can be sustained regardless of the batch size. These observations lead us to a conclusion that in the presence of those two applications, the workload offloading to the integrated graphics card is a must, as it not only achieves top performance but it also leaves the CPU and the discrete GPU idle, which either promotes the energy efficiency of the system as a whole or provides room for the execution of at least one computation-intensive application, like AES, without sacrificing performance.

## VI. REAL-TIME SCHEDULING

Our scheduling system consists of two phases, the offline analysis part, and the online adaptive scheduling.

### A. Offline Analysis

In the first phase, we implement an offline analysis tool that creates all application combinations possible, which then tests on every device combination and gathers the resulted performance statistics. By doing so, all possible configurations are systematically characterized; these can be used later at real-time for deciding the most appropriate based on the corresponding conditions (see Section VI-B.

The user of our system is able to specify the number of applications to run, the number of network interfaces per application (at least one) and the training time of the offline analysis (how much time our system will spend on each configuration profiling). For instance, let us assume that a user wants to run application A and application B on a single processor machine with only an integrated GPU available (no discrete GPUs or any other co-processors or accelerators). The user also specifies that the system needs to poll all four network interfaces and that each application of a configuration will process twenty batches of packets. Our system then takes charge of the rest. First, it creates every possible configuration and maps the corresponding applications to each. In this example case, there are two physical devices and two different applications, so the combinations of applications-to-devices are the following four: (i) both applications will run on the CPU, (ii) both applications will run on the integrated GPU, (iii) A will run on the CPU, while B will run on the integrated GPU and (iv) B will run on the CPU, while A will run on the integrated GPU. Then, the system introduces batching to ensure that there are enough configurations to effectively measure both latency and throughput. There are 7 different batch sizes (1K, 2K, 4K, 8K, 16K, 32K, 64K), so for each application-to-device combination there are seven different system configurations, which leads to a total of 28 configurations. Then, our system activate the network interfaces and assign two of them to each application. Finally our system activates the first configuration, waits until it has processed twenty batches of packets, stores the details of the configuration alongside with its statistics and performance in a red-black tree (as it guarantees efficiency of the basic data structure operations), deactivates the configuration and then repeat the same process for every configuration that is available to run.

### B. Live Scheduling

At this point our system has populated a tree with the performance results of each configuration. The user specifies a policy (detailed information about the policies in the following subsection) that is based on his needs and our system initially searches the populated tree, picks the best performing configuration with respect to the user defined policy and processes incoming packets from the network. As traffic rate changes, we want our system to adapt and be able to keep processing the input data as efficient as possible. In order to achieve this, our system periodically checks if the best configuration is still active and if it is not, it activates it. There is a monitor thread responsible for managing whether the best configuration is active and its complete functionality is thoroughly explained below.

*a) Monitor:* Our system needs an extra thread to be able to keep track of the active configuration and deciding whether to keep running under the current configuration or changing to a new best. To save system resources, we decide not to spawn an extra thread. Instead, we use the main thread of our system, which is idle after performing the offline analysis, to run as the monitor. The implementation is fairly simple, we (i) setup a consistent alarm using POSIX reliable signals, (ii) implement a handler function that is called every time the alarm timer expires, (iii) the user is able to either explicitly specify the timer of use the system default.

At first, there is a dispatcher that calls the appropriate policy function based on the user's choice to decide if the active configuration is still performing better that any other configuration. This decision is made by comparing the performance statistics between the last and the second to last call of the handler function to the performance statistics of the other configurations. If the current configuration does not perform optimally, then the system deactivates it and activates the new-best based on the results of the aforementioned comparison. Next step of the handler function is to update the performance statistics of the current active configuration to match the most recent performance statistics of the system. This update is necessary for our system as it keeps training itself over time and can successfully adapt to any traffic changes regardless the size of the variance. The final step is to update the timer that indicates when the scheduler will eventually shut down, if it is set by the user. This is mainly for testing purposes, generally no timer is set so the scheduler will not stop its execution.

*b) Policy-on-the-fly:* Although our system is highly adaptive and can detect traffic fluctuations, we do recognize that the user may need to update the way the system performs by applying a different policy under which the system will decide the workload distribution without re-initializing the whole system. To achieve that we implement a way for the user to be able to update the policy real-time. To do so, we create a custom signal handler for catching software generated interrupts. From that point on, a different policy can be applied. Our system detects the change and distributes the workload accordingly.

TABLE I
PERFORMANCE CHARACTERIZATION OF THE DISCRETE GPU (NVIDIA GTX 1080 TI), THE CPU (I7-8700K) AND THE INTEGRATED GPU (UHD GRAPHICS 630) USING THE DPI APPLICATION IN CONJUNCTION WITH DIFFERENT COMBINATIONS OF CO-WORKERS.

| Device | Application | Buffer size | Co-workers | Co-worker | Performance per Kernel | | | | Aggregated performance | | |
|---|---|---|---|---|---|---|---|---|---|---|---|
| | | | | | msec | Mpps | Gbps | Slow-down | msec | Mpps | Gbps |
| GTX1080Ti | DPI | 1024 | 1 | MD5 | 1.0 | 1.099 | 12.9 | 14.6% | 1.9 | 2.196 | 25.8 |
| | | | 3 | | 1.7 | 0.588 | 6.9 | 54.3% | 7.0 | 2.351 | 27.6 |
| | | 4096 | 1 | | 4.0 | 1.019 | 12.0 | 33.3% | 8.0 | 2.038 | 23.9 |
| | | | 3 | | 7.0 | 0.587 | 6.9 | 61.7% | 27.9 | 2.346 | 27.6 |
| | | 16384 | 1 | | 22.7 | 0.722 | 8.5 | 49.4% | 43.7 | 1.503 | 17.7 |
| | | | 3 | | 41.0 | 0.399 | 4.7 | 72.0% | 155.9 | 1.684 | 19.8 |
| | | 1024 | 1 | AES | 0.9 | 1.133 | 13.3 | 11.9% | 1.8 | 2.270 | 26.7 |
| | | | 3 | | 1.7 | 0.588 | 6.9 | 54.3% | 7.0 | 2.351 | 27.6 |
| | | 4096 | 1 | | 3.7 | 1.097 | 12.9 | 28.3% | 7.5 | 2.195 | 25.8 |
| | | | 3 | | 7.0 | 0.586 | 6.9 | 61.7% | 27.9 | 2.346 | 27.6 |
| | | 16384 | 1 | | 26.9 | 0.610 | 7.2 | 57.1% | 53.8 | 1.219 | 14.3 |
| | | | 3 | | 39.2 | 0.418 | 4.9 | 70.8% | 154.1 | 1.702 | 20.0 |
| | | 1024 | 1 | DPI | 0.9 | 1.177 | 13.8 | 8.6% | 1.7 | 2.353 | 27.6 |
| | | | 3 | | 1.7 | 0.588 | 6.9 | 54.3% | 7.0 | 2.352 | 27.6 |
| | | 4096 | 1 | | 3.7 | 1.112 | 13.1 | 27.2% | 7.4 | 2.223 | 26.1 |
| | | | 3 | | 7.0 | 0.587 | 6.9 | 61.7% | 27.9 | 2.347 | 27.6 |
| | | 16384 | 1 | | 21.5 | 0.761 | 8.9 | 47.0% | 43.1 | 1.521 | 17.9 |
| | | | 3 | | 36.2 | 0.453 | 5.3 | 68.5% | 145.4 | 1.803 | 21.2 |
| i7-8700K | DPI | 1024 | 1 | MD5 | 0.9 | 1.204 | 14.1 | 45.8% | 1.8 | 2.307 | 27.1 |
| | | | 3 | | 1.8 | 0.567 | 6.7 | 74.2% | 7.1 | 2.339 | 27.5 |
| | | 4096 | 1 | | 3.3 | 1.230 | 14.5 | 42.9% | 7.1 | 2.318 | 27.2 |
| | | | 3 | | 7.2 | 0.567 | 6.7 | 73.6% | 28.3 | 2.323 | 27.3 |
| | | 16384 | 1 | | 13.2 | 1.245 | 14.6 | 42.7% | 30.5 | 2.188 | 25.7 |
| | | | 3 | | 28.9 | 0.567 | 6.7 | 73.7% | 115.9 | 2.265 | 26.6 |
| | | 1024 | 1 | AES | 0.6 | 1.586 | 18.6 | 28.5% | 2.1 | 2.314 | 27.2 |
| | | | 3 | | 1.5 | 0.689 | 8.1 | 68.8% | 9.4 | 1.857 | 21.8 |
| | | 4096 | 1 | | 2.9 | 1.404 | 16.5 | 35.0% | 8.3 | 2.171 | 25.5 |
| | | | 3 | | 6.7 | 0.610 | 7.2 | 71.7% | 38.4 | 1.776 | 20.9 |
| | | 16384 | 1 | | 12.6 | 1.299 | 15.3 | 40.0% | 34.8 | 2.039 | 24.0 |
| | | | 3 | | 29.8 | 0.549 | 6.5 | 74.5% | 155.5 | 1.727 | 20.3 |
| | | 1024 | 1 | DPI | 0.8 | 1.274 | 15.0 | 42.3% | 1.8 | 2.290 | 26.9 |
| | | | 3 | | 1.7 | 0.604 | 7.1 | 72.7% | 7.1 | 2.327 | 27.3 |
| | | 4096 | 1 | | 3.3 | 1.230 | 14.5 | 42.9% | 7.1 | 2.312 | 27.2 |
| | | | 3 | | 6.8 | 0.601 | 7.1 | 72.0% | 28.5 | 2.306 | 27.1 |
| | | 16384 | 1 | | 13.0 | 1.261 | 14.8 | 41.2% | 30.9 | 2.178 | 25.6 |
| | | | 3 | | 28.9 | 0.567 | 6.7 | 73.7% | 120.5 | 2.180 | 25.6 |
| UHDGraphics | DPI | 1024 | 1 | MD5 | 0.9 | 1.144 | 13.4 | 47.7% | 1.8 | 2.288 | 26.9 |
| | | | 3 | | 1.8 | 0.583 | 6.9 | 73.0% | 7.0 | 2.333 | 27.4 |
| | | 4096 | 1 | | 3.5 | 1.176 | 13.8 | 50.0% | 7.0 | 2.352 | 27.6 |
| | | | 3 | | 7.0 | 0.588 | 6.9 | 75.0% | 27.9 | 2.351 | 27.6 |
| | | 16384 | 1 | | 13.9 | 1.177 | 13.8 | 50.2% | 27.8 | 2.353 | 27.7 |
| | | | 3 | | 27.9 | 0.588 | 6.9 | 75.1% | 111.4 | 2.352 | 27.6 |
| | | 1024 | 1 | AES | 1.3 | 0.803 | 9.4 | 63.3% | 2.6 | 1.607 | 18.9 |
| | | | 3 | | 3.2 | 0.321 | 3.8 | 85.2% | 12.8 | 1.283 | 15.1 |
| | | 4096 | 1 | | 4.3 | 0.963 | 11.3 | 59.1% | 8.5 | 1.924 | 22.6 |
| | | | 3 | | 11.2 | 0.367 | 4.3 | 84.4% | 44.7 | 1.467 | 17.2 |
| | | 16384 | 1 | | 16.5 | 0.997 | 11.7 | 57.8% | 33.0 | 1.990 | 23.4 |
| | | | 3 | | 42.1 | 0.389 | 4.6 | 83.4% | 169.2 | 1.549 | 18.2 |
| | | 1024 | 1 | DPI | 0.9 | 1.129 | 13.3 | 48.0% | 1.8 | 2.257 | 26.5 |
| | | | 3 | | 1.8 | 0.576 | 6.8 | 73.4% | 7.1 | 2.302 | 27.1 |
| | | 4096 | 1 | | 3.5 | 1.177 | 13.8 | 50.0% | 7.0 | 2.353 | 27.6 |
| | | | 3 | | 7.0 | 0.588 | 6.9 | 75.0% | 27.9 | 2.352 | 27.6 |
| | | 16384 | 1 | | 13.9 | 1.177 | 13.8 | 50.2% | 27.9 | 2.353 | 27.6 |
| | | | 3 | | 27.9 | 0.588 | 6.9 | 75.1% | 111.4 | 2.352 | 27.6 |

### C. Policies

We define and implement four different policy algorithms which are enough to fulfil the different needs of a potential user of our system. There is a short description of each predefined policy below. Furthermore, if a user has special requirements, it is fairly easy to define his own policy and use it. We also shortly describe how this can be achieved.

*a) Maximize Throughput:* The first policy we implement is the maximum throughput policy. Our scheduler chooses the configuration that gives the maximum aggregated rate (in Gbits per second) at which the target devices process input data, regardless of the other performance metrics. This is the best configuration when the user wants to maximize the throughput of the system but comes with the cost of greatly increasing the latency due to generally large-sized batches.

*b) Minimize Latency:* The second policy is completely opposing to the first, as the goal is to minimize the end to end latency. In other words the user wants each packet to be processed as soon as it is captured. This kind of policy applies to latency sensitive applications that provide real-time processing as a feature.

TABLE II
PERFORMANCE CHARACTERIZATION OF THE DISCRETE GPU (NVIDIA GTX 1080 TI), THE CPU (I7-8700K) AND THE INTEGRATED GPU (UHD GRAPHICS 630) USING THE AES APPLICATION IN CONJUNCTION WITH DIFFERENT COMBINATIONS OF CO-WORKERS.

| Device | Application | Buffer size | Co-workers | Co-worker | Performance per Kernel | | | | Aggregated performance | | |
|---|---|---|---|---|---|---|---|---|---|---|---|
| | | | | | msec | Mpps | Gbps | Slow-down | msec | Mpps | Gbps |
| GTX1080Ti | AES | 1024 | 1 | MD5 | 0.9 | 1.176 | 13.8 | 26.2% | 1.7 | 2.352 | 27.6 |
| | | | 3 | | 1.7 | 0.588 | 6.9 | 63.1% | 7.0 | 2.351 | 27.6 |
| | | 4096 | 1 | | 4.0 | 1.032 | 12.1 | 34.2% | 7.9 | 2.064 | 24.3 |
| | | | 3 | | 7.0 | 0.586 | 6.9 | 62.5% | 27.9 | 2.347 | 27.6 |
| | | 16384 | 1 | | 22.9 | 0.715 | 8.4 | 49.4% | 43.6 | 1.508 | 17.7 |
| | | | 3 | | 40.9 | 0.400 | 4.7 | 71.7% | 158.6 | 1.653 | 19.4 |
| | | 1024 | 1 | AES | 0.9 | 1.177 | 13.8 | 26.2% | 1.7 | 2.353 | 27.6 |
| | | | 3 | | 1.7 | 0.588 | 6.9 | 63.1% | 7.0 | 2.352 | 27.6 |
| | | 4096 | 1 | | 3.7 | 1.121 | 13.2 | 28.3% | 7.3 | 2.242 | 26.3 |
| | | | 3 | | 7.0 | 0.587 | 6.9 | 62.5% | 27.9 | 2.347 | 27.6 |
| | | 16384 | 1 | | 21.6 | 0.758 | 8.9 | 46.4% | 43.2 | 1.516 | 17.8 |
| | | | 3 | | 36.2 | 0.452 | 5.3 | 68.1% | 145.1 | 1.807 | 21.2 |
| | | 1024 | 1 | DPI | 0.9 | 1.128 | 13.3 | 28.9% | 1.9 | 2.249 | 26.4 |
| | | | 3 | | 1.7 | 0.588 | 6.9 | 63.1% | 7.0 | 2.350 | 27.6 |
| | | 4096 | 1 | | 3.6 | 1.128 | 13.3 | 27.7% | 7.3 | 2.257 | 26.5 |
| | | | 3 | | 7.0 | 0.587 | 6.9 | 62.5% | 27.9 | 2.347 | 27.6 |
| | | 16384 | 1 | | 26.0 | 0.629 | 7.4 | 55.4% | 52.1 | 1.259 | 14.8 |
| | | | 3 | | 39.7 | 0.412 | 4.8 | 71.1% | 158.6 | 1.653 | 19.4 |
| i7-8700K | AES | 1024 | 1 | MD5 | 1.2 | 0.880 | 10.3 | 41.8% | 1.9 | 2.331 | 27.4 |
| | | | 3 | | 2.3 | 0.438 | 5.1 | 71.2% | 7.3 | 2.320 | 27.3 |
| | | 4096 | 1 | | 4.8 | 0.854 | 10.0 | 45.4% | 7.8 | 2.200 | 25.8 |
| | | | 3 | | 9.4 | 0.439 | 5.2 | 71.6% | 29.6 | 2.267 | 26.6 |
| | | 16384 | 1 | | 20.9 | 0.783 | 9.2 | 49.2% | 33.5 | 2.085 | 24.5 |
| | | | 3 | | 35.3 | 0.464 | 5.4 | 70.2% | 117.3 | 2.265 | 26.6 |
| | | 1024 | 1 | AES | 1.2 | 0.852 | 10.0 | 43.5% | 2.6 | 1.611 | 18.9 |
| | | | 3 | | 2.5 | 0.406 | 4.8 | 72.9% | 10.7 | 1.559 | 18.3 |
| | | 4096 | 1 | | 5.0 | 0.827 | 9.7 | 46.7% | 10.4 | 1.579 | 18.6 |
| | | | 3 | | 10.5 | 0.391 | 4.6 | 74.9% | 43.2 | 1.523 | 17.9 |
| | | 16384 | 1 | | 20.0 | 0.821 | 9.6 | 47.0% | 42.0 | 1.563 | 18.4 |
| | | | 3 | | 42.2 | 0.389 | 4.6 | 74.6% | 174.7 | 1.506 | 17.7 |
| | | 1024 | 1 | DPI | 1.3 | 0.807 | 9.5 | 46.3% | 2.0 | 2.322 | 27.3 |
| | | | 3 | | 2.5 | 0.407 | 4.8 | 72.9% | 7.4 | 2.286 | 26.9 |
| | | 4096 | 1 | | 5.3 | 0.771 | 9.1 | 50.3% | 8.3 | 2.164 | 25.4 |
| | | | 3 | | 10.1 | 0.407 | 4.8 | 73.8% | 30.5 | 2.215 | 26.0 |
| | | 16384 | 1 | | 21.4 | 0.764 | 9.0 | 50.3% | 35.1 | 1.966 | 23.1 |
| | | | 3 | | 38.3 | 0.428 | 5.0 | 72.4% | 126.8 | 2.104 | 24.7 |
| UHDGraphics | AES | 1024 | 1 | MD5 | 1.3 | 0.821 | 9.6 | 10.3% | 2.5 | 1.641 | 19.3 |
| | | | 3 | | 1.8 | 0.560 | 6.6 | 38.3% | 7.3 | 2.241 | 26.3 |
| | | 4096 | 1 | | 4.4 | 0.931 | 10.9 | 6.0% | 8.8 | 1.862 | 21.9 |
| | | | 3 | | 7.1 | 0.576 | 6.8 | 41.4% | 28.5 | 2.302 | 27.0 |
| | | 16384 | 1 | | 16.6 | 0.986 | 11.6 | 7.2% | 33.2 | 1.972 | 23.2 |
| | | | 3 | | 28.4 | 0.576 | 6.8 | 45.6% | 113.8 | 2.303 | 27.1 |
| | | 1024 | 1 | AES | 2.0 | 0.504 | 5.9 | 44.9% | 4.1 | 1.008 | 11.8 |
| | | | 3 | | 3.9 | 0.265 | 3.1 | 71.0% | 15.5 | 1.059 | 12.4 |
| | | 4096 | 1 | | 7.3 | 0.560 | 6.6 | 43.1% | 14.6 | 1.120 | 13.2 |
| | | | 3 | | 13.9 | 0.296 | 3.5 | 69.8% | 55.5 | 1.181 | 13.9 |
| | | 16384 | 1 | | 27.7 | 0.590 | 6.9 | 44.8% | 55.6 | 1.178 | 13.8 |
| | | | 3 | | 52.9 | 0.310 | 3.6 | 71.2% | 212.3 | 1.235 | 14.5 |
| | | 1024 | 1 | DPI | 1.3 | 0.801 | 9.4 | 12.1% | 2.6 | 1.601 | 18.8 |
| | | | 3 | | 1.9 | 0.552 | 6.5 | 39.3% | 7.4 | 2.208 | 25.9 |
| | | 4096 | 1 | | 4.3 | 0.963 | 11.3 | 2.6% | 8.5 | 1.926 | 22.6 |
| | | | 3 | | 7.1 | 0.575 | 6.8 | 41.4% | 28.5 | 2.300 | 27.0 |
| | | 16384 | 1 | | 16.5 | 0.993 | 11.7 | 6.4% | 33.0 | 1.992 | 23.4 |
| | | | 3 | | 28.5 | 0.576 | 6.8 | 45.6% | 113.6 | 2.307 | 27.1 |

*c) Minimize Energy Consumption:* The purpose of the next policy is to reduce energy consumption as much as possible. As a consequence the aggregated throughput of the system is considerably low. When only one target device suffices and the rest are idle, the latency of the system is not a major problem.

*d) Process 10 Gbps:* This one is a special case policy that ensures that the system will process packets at a rate as close to 10 Gbits per second as possible. It can be useful when there are power constraints but not as strict as in the previous policy. In this case the user does not want the system to reduce its energy consumption by devastating its overall performance (latency and/or throughput).

*e) User Defined Policies:* Apart from the policies we provide, it is fairly easy for a user to create a custom one and run our system using that instead of our policies. There are 3 main steps in order to implement and attach a new user defined policy to our system, which are the following: (i) implement a function with the following signature `void userDefinedPolicy(void);` This is the main function that checks if the best configuration is active and if not, it activates it. The checking takes into account the performance metric or the combination of more than one performance metrics that the user wants to minimize or maximize. (ii) inform the policy function dispatcher about the existence of the newly created function and (iii) recompile the system.

TABLE III
PERFORMANCE CHARACTERIZATION OF THE DISCRETE GPU (NVIDIA GTX 1080 TI), THE CPU (I7-8700K) AND THE INTEGRATED GPU (UHD
GRAPHICS 630) USING THE MD5 APPLICATION IN CONJUNCTION WITH DIFFERENT COMBINATIONS OF CO-WORKERS.

| Device | Application | Buffer size | Co-workers | Co-worker | Performance per Kernel | | | | Aggregated performance | | |
|---|---|---|---|---|---|---|---|---|---|---|---|
| | | | | | msec | Mpps | Gbps | Slow-down | msec | Mpps | Gbps |
| GTX1080Ti | MD5 | 1024 | 1 | MD5 | 0.9 | 1.176 | 13.8 | 37.0% | 1.7 | 2.353 | 27.6 |
| | | | 3 | | 1.7 | 0.558 | 6.9 | 68.5% | 7.0 | 2.352 | 27.6 |
| | | 4096 | 1 | | 3.5 | 1.159 | 13.6 | 32.3% | 7.1 | 2.320 | 27.3 |
| | | | 3 | | 7.0 | 0.587 | 6.9 | 65.7% | 27.9 | 2.348 | 27.6 |
| | | 16384 | 1 | | 21.2 | 0.772 | 9.1 | 48.9% | 42.4 | 1.545 | 18.2 |
| | | | 3 | | 36.2 | 0.453 | 5.3 | 70.2% | 144.9 | 1.810 | 21.3 |
| | | 1024 | 1 | AES | 0.9 | 1.176 | 13.8 | 37.0% | 1.7 | 2.353 | 27.6 |
| | | | 3 | | 1.7 | 0.588 | 6.9 | 68.5% | 7.0 | 2.351 | 27.6 |
| | | 4096 | 1 | | 4.0 | 1.034 | 12.2 | 39.3% | 7.9 | 2.066 | 24.3 |
| | | | 3 | | 7.0 | 0.587 | 6.9 | 65.7% | 27.9 | 2.348 | 27.6 |
| | | 16384 | 1 | | 21.6 | 0.759 | 8.9 | 50.0% | 43.7 | 1.499 | 17.6 |
| | | | 3 | | 38.0 | 0.431 | 5.1 | 71.3% | 160.1 | 1.639 | 19.3 |
| | | 1024 | 1 | DPI | 1.0 | 1.082 | 12.7 | 42.0% | 2.0 | 2.168 | 25.5 |
| | | | 3 | | 1.7 | 0.588 | 6.9 | 68.5% | 7.0 | 2.351 | 27.6 |
| | | 4096 | 1 | | 4.0 | 1.020 | 12.0 | 40.3% | 8.0 | 2.038 | 23.9 |
| | | | 3 | | 7.0 | 0.587 | 6.9 | 65.7% | 27.9 | 2.347 | 27.6 |
| | | 16384 | 1 | | 20.9 | 0.783 | 9.2 | 48.3% | 43.4 | 1.512 | 17.8 |
| | | | 3 | | 38.9 | 0.421 | 4.9 | 72.5% | 167.2 | 1.570 | 18.5 |
| i7-8700K | MD5 | 1024 | 1 | MD5 | 0.9 | 1.197 | 14.1 | 47.2% | 1.8 | 2.335 | 27.4 |
| | | | 3 | | 1.7 | 0.597 | 7.0 | 73.8% | 7.1 | 2.344 | 27.5 |
| | | 4096 | 1 | | 3.5 | 1.174 | 13.8 | 49.5% | 7.0 | 2.334 | 27.4 |
| | | | 3 | | 6.7 | 0.610 | 7.2 | 73.6% | 28.1 | 2.335 | 27.4 |
| | | 16384 | 1 | | 13.9 | 1.180 | 13.9 | 42.8% | 28.4 | 2.306 | 27.1 |
| | | | 3 | | 26.7 | 0.614 | 7.2 | 70.4% | 113.9 | 2.307 | 27.1 |
| | | 1024 | 1 | AES | 0.7 | 1.519 | 17.8 | 33.3% | 2.0 | 2.317 | 27.2 |
| | | | 3 | | 1.4 | 0.756 | 8.9 | 66.7% | 9.1 | 1.946 | 22.9 |
| | | 4096 | 1 | | 3.0 | 1.379 | 16.2 | 40.7% | 7.9 | 2.210 | 26.0 |
| | | | 3 | | 6.2 | 0.665 | 7.8 | 71.4% | 37.1 | 1.859 | 21.8 |
| | | 16384 | 1 | | 12.8 | 1.283 | 15.1 | 37.9% | 33.7 | 2.066 | 24.3 |
| | | | 3 | | 27.0 | 0.607 | 7.1 | 70.8% | 149.7 | 1.810 | 21.3 |
| | | 1024 | 1 | DPI | 0.9 | 1.176 | 13.8 | 48.3% | 1.8 | 2.307 | 27.1 |
| | | | 3 | | 1.7 | 0.594 | 7.0 | 73.8% | 7.1 | 2.329 | 27.4 |
| | | 4096 | 1 | | 3.3 | 1.230 | 14.5 | 46.9% | 7.1 | 2.317 | 27.2 |
| | | | 3 | | 7.6 | 0.541 | 6.4 | 76.6% | 28.5 | 2.306 | 27.1 |
| | | 16384 | 1 | | 14.5 | 1.130 | 13.3 | 45.3% | 30.2 | 2.172 | 25.5 |
| | | | 3 | | 28.4 | 0.578 | 6.8 | 72.0% | 117.5 | 2.239 | 26.3 |
| UHDGraphics | MD5 | 1024 | 1 | MD5 | 0.9 | 1.158 | 13.6 | 49.3% | 1.8 | 2.316 | 27.2 |
| | | | 3 | | 1.8 | 0.583 | 6.9 | 74.3% | 7.0 | 2.333 | 27.4 |
| | | 4096 | 1 | | 3.5 | 1.177 | 13.8 | 50.2% | 7.0 | 2.353 | 27.6 |
| | | | 3 | | 7.0 | 0.588 | 6.9 | 75.1% | 27.9 | 2.352 | 27.6 |
| | | 16384 | 1 | | 13.9 | 1.177 | 13.8 | 50.2% | 27.9 | 2.353 | 27.6 |
| | | | 3 | | 27.9 | 0.588 | 6.9 | 75.1% | 111.4 | 2.352 | 27.6 |
| | | 1024 | 1 | AES | 1.2 | 0.824 | 9.7 | 63.8% | 2.5 | 1.649 | 19.4 |
| | | | 3 | | 3.1 | 0.327 | 3.8 | 85.8% | 12.5 | 1.309 | 15.4 |
| | | 4096 | 1 | | 4.3 | 0.956 | 11.2 | 59.6% | 8.6 | 1.912 | 22.5 |
| | | | 3 | | 11.3 | 0.364 | 4.3 | 84.5% | 45.1 | 1.453 | 17.1 |
| | | 16384 | 1 | | 16.6 | 0.986 | 11.6 | 58.1% | 33.2 | 1.973 | 23.2 |
| | | | 3 | | 42.3 | 0.387 | 4.6 | 83.4% | 169.9 | 1.543 | 18.1 |
| | | 1024 | 1 | DPI | 0.9 | 1.153 | 13.5 | 49.6% | 1.8 | 2.306 | 27.1 |
| | | | 3 | | 1.8 | 0.577 | 6.8 | 74.6% | 7.1 | 2.309 | 27.1 |
| | | 4096 | 1 | | 3.5 | 1.176 | 13.8 | 50.2% | 7.0 | 2.352 | 27.6 |
| | | | 3 | | 7.0 | 0.588 | 6.9 | 75.1% | 27.9 | 2.351 | 27.6 |
| | | 16384 | 1 | | 13.9 | 1.176 | 13.8 | 50.2% | 27.8 | 2.353 | 27.7 |
| | | | 3 | | 27.9 | 0.588 | 6.9 | 75.1% | 111.4 | 2.352 | 27.6 |

## VII. EVALUATION

In this section, we evaluate the performance of our scheduling algorithm, using packet processing applications described in Section III-D. Specifically, due to space constraints, we evaluate our scheduler using two diverse applications, AES and DPI. In Figures 5(a), 5(b), 5(c) and 5(d) we present the input rates and achieved throughput, the power consumption of the device combination made by the scheduler, in dynamic conditions: i.e. (i) fluctuating incoming network traffic rate and (ii) policy change on-the-fly .

Firstly, regarding the fluctuating network traffic rate experiment, we use an energy-critical policy to handle all input traffic at maximum energy efficiency. The traffic rate is low enough for a single device to cope with it, which results to significantly low power consumption. This is not the case in the second experiment, where we seek the highest possible throughput before aggressively switching to an energy-efficient policy. We note that the power consumption is divided into two categories: the power consumption of the (i) GTX 1080 Ti (*"GPU"*) and (ii) Intel i7-8700K die, which also includes the power consumption of the Intel UHD Graphics 630 (*"CPU die"*). This restriction goes hand in hand with the existing hardware limitations, as it is not possible to distinguish the power consumption of the different parts of the same CPU die using software. For comparison, we also display with a straight gray line the maximum power consumption when both devices are exhaustively used simultaneously. Additionally, we provide the processing latency with a solid black line. The observed variability in latency is the result of dynamic scheduler decisions regarding the batching and device selection. The input traffic is composed by 1514-byte TCP packets.

Overall, our scheduler is capable to adapt to a highly diverse computational demand among different applications, producing live decisions that aim to maintain the maximum energy efficiency during the total execution (besides the requested performance policy). Additionally, the scheduler can sustain high throughput and avoids selecting device combinations that lead to excessive latency.

*a) Throughput:* As we observe through our experiments, our proposed scheduler is able to choose the configuration that keeps the selected device under the processing capacity which is required to process the incoming traffic for each application. Specifically, when traffic rate is constantly high and the specified policy is to hit the maximum throughput, the system is able to process the input traffic at a rate of almost 20 Gbps, if only a single application is active and at a rate of almost 30 Gbps in the scenario of two active applications, as shown in Figures 5(c) and 5(d) respectively (between times 0 and 15 seconds). On the other hand, when the traffic rate is variable, the scheduling schema we propose manages to cope with up to 10 Gbps input traffic rate per application by activating a single device and interleaving applications, if more that one exists, as shown in Figures 5(a) and 5(b) (at times 0 to 20). When locating changes in the traffic rate, such as the increase from 20 to 40 Gbps (Figure 5(b)), a second device (in this case the GTX 1080 Ti) is enabled to increase the computational capacity of the system. An interesting area is located at times 20 to 30 of Figure 5(a) when the discrete graphics card is activated but is immediately deactivated, as the monitoring reveals that only the presence of the integrated graphics card can still cope with the incoming traffic. The GTX 1080 Ti is only re-activated when the traffic rate is doubled to 40 Gbps.

*b) Energy Efficiency:* The more working devices lead to higher power consumption and vice versa. That is the reason behind the decision to design a system that constantly adapts by activating the least possible number of devices needed to meet the requested requirements. In Figures 5(c) and 5(d) at the 15-second mark the system switches from highest throughput to lowest power consumption policy, thus is shuts down the power-hungry GTX 1080 Ti, resulting in a severe drop in the overall energy usage. It is also clear from the results of variable traffic rate experiments (Figures 5(a) and 5(b)) that only when a severe increase in the traffic rate occurs and more computational capacity is needed, the system activates an extra device (the GTX 1080 Ti in this case) at the cost of greater energy expenditure.

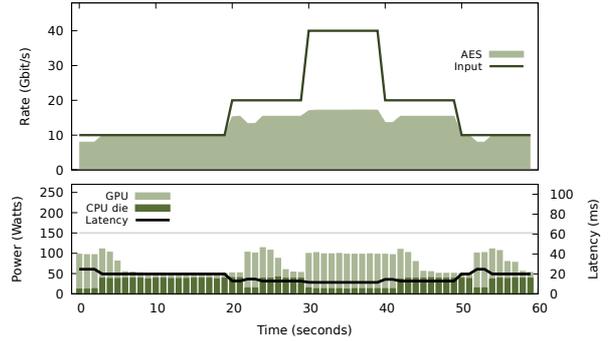
(a) Fluctuating input rate while executing AES.

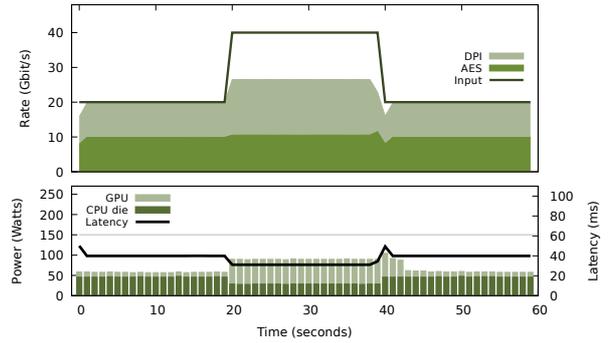
(b) Fluctuating input rate while executing AES and DPI.

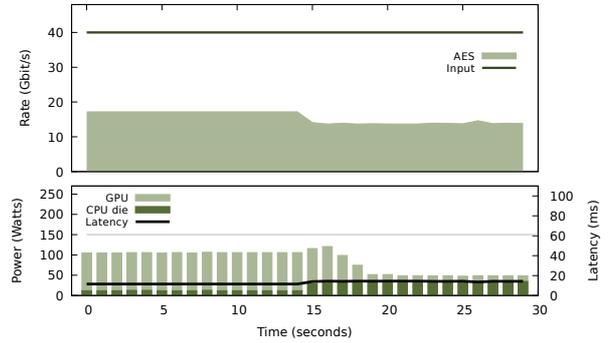
(c) Policy change (*maximize throughput* to *minimize power consumption*) while executing AES.

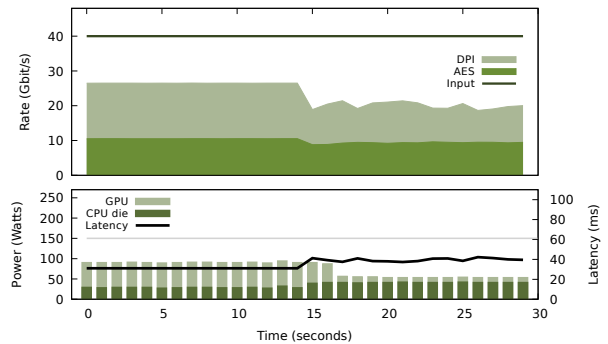
(d) Policy change (*maximize throughput* to *minimize power consumption*) while executing AES and DPI.

Fig. 5. Automatic device configuration selection under different conditions (i.e. incoming traffic rate fluctuations and policy change) for different combinations of applications.

TABLE IV
ANALYSIS OF PREVIOUS WORKS. PYTHIA FILLS THE GAP BY PROPOSING A SCHEDULING ALGORITHM THAT DISTRIBUTES THE EXECUTION OF MULTIPLE NETWORK PACKET PROCESSING APPLICATIONS ACROSS MULTIPLE HETEROGENEOUS PROCESSING UNITS, IN ORDER TO ACHIEVE BETTER PERFORMANCE.

|  | Target Architecture | | Application support | |
| --- | --- | --- | --- | --- |
| Related work | Single Dev | Multi Devs | Single App | Multi Apps |
| *GASPP* [38] | ✓ | – | ✓ | – |
| *Dobrescu et al.* [11] | ✓ | – | ✓ | ✓ |
| *Papadogiannaki et al.* [31] | ✓ | ✓ | ✓ | – |
| **Pythia** | ✓ | ✓ | ✓ | ✓ |

*c) Latency:* As observed, increasing the batch size results to higher throughput rates; however, at the cost of increased latency, especially in the case of the discrete GPU. However, we try to minimize latency up to a point where no interference with the requested policy occurs. For example, even when the goal is to maximize the overall throughput of the system, like in Figure 5(b), during the second 20-seconds interval, latency remains considerably low despite the fact that the discrete GPU is active as the traffic characteristics demands so. The reason behind this is not only the presence of an extra device, but mainly because the system does recognize that an even larger batch size would not result in extra performance gains.

*d) Traditional Performance Metrics:* Besides the above performance metrics, we also take account other important performance metrics in the domain of software network packet processing, i.e. the packet reordering and packet loss. Reordering can occur when packets that belong in a single flow get split and distributed to different devices. We solve this issue by ensuring the following condition; never process different parts of the same flow in more than one different devices. Furthermore, packet loss can occur in the case of slow device switches due to a scheduling decision that aims to adapt to reduced or increased traffic rate. Our scheduler though, is able to promptly adapt to changes in less than 60ms, resulting to no packet losses, even in cases where the input traffic rate increases from 20Gbps to 40Gbps.

## VIII. RELATED WORK

Recently, GPUs have become very popular due to a substantial performance boost that provide to many individual network traffic inspection applications, such as including intrusion detection [34], [37], [39], [40], [40], cryptography [18], and IP routing [17]. In addition, there have been proposed several programmable network traffic processing frameworks, such as Snap [36] and GASPP [38], that manage to simplify the development of GPU-accelerated network traffic processing applications. Unlike these works, we focus on building a software network packet processing framework that combines different, heterogeneous, processing devices and quantify the problems that arise through their concurrent utilization. By effectively mapping computations to heterogeneous devices, in an automated way, we provide more efficient execution in terms of throughput, latency and power consumption.

Load-balancing systems support applications with multiple concurrent kernels [10], [16], [35], yet, these approaches are not evaluated in dynamic environments using fluctuating inputs, such as network traffic. Other approaches rely heavily on manual intervention by the programmer [24], [28]. Approaches to load-balance a single computational kernel include [8], [21], [25], [27]. The simplest approach [27] target homogeneous GPUs and require no training as they use a fixed work partition [21]. Wang and Ren propose a distribution method on a CPU-GPU heterogeneous system that tries a large number of different work distributions to find the most efficient [41]. Another work proposes treating the CPU as the primary processor, since it offers low latency, and offloading processing tasks to accelerators only when they result in throughput benefit [20]. Unlike this work, we take into account the power characteristics of each computational device, and also target more heterogeneous processors (like integrated GPUs that have high on-chip dependencies with the CPU). In APUNet, the authors take advantage of the APU architecture to accelerate packet processing applications [15]. Other works propose adaptive scheduling approaches for heterogeneous hardware architectures tailored for network packet processing applications [22], [31]. The scheduler uses performance policies (i.e. high throughput, low latency or low power consumption) to determine the appropriate combination of devices for efficient execution. In this work, we extend the former solution enabling the concurrent execution of different network packet processing applications across heterogeneous hardware architectures.

Other approaches require a series of small execution trials to determine the performance [8], [25]. Still, these approaches have been designed for applications that take as input constant streaming data and, thus, they struggle to adapt when the input data stream varies. Thus, such solutions are tough to be applied to network processing applications in which the heterogeneity of *(i)* the hardware, *(ii)* the applications, and *(iii)* the traffic vastly affect the overall efficiency in terms of performance and power consumption. To that end, our proposed scheduling algorithm has been designed to explicitly account for these conditions. Furthermore, there is ongoing work on providing performance predictability [11] and fair queuing [13] when running a diverse set of applications that contend for shared hardware resources. Finally, there is also work on packet routing [29] that draws power proportional to the traffic load. The main difference with our work, is that they focus solely on homogeneous processing cores; instead, we present a system that utilizes efficiently a diverse set of devices.

## IX. CONCLUSIONS

In this paper we propose an adaptive and highly dynamic scheduling solution tailored explicitly for network packet processing applications. Our approach enables real-time application multiplexing across heterogeneous and asymmetric architectures that can be found on commodity, off-the-shelf hardware setups. In this work, we manage to improve the overall efficiency of the tested applications, since our scheduler is able to choose the configuration that results to the optimal performance each time relatively to the current state, responding quickly either to network fluctuations or system changes. As part of our future work, we plan to apply machine learning techniques to predict the most efficient configuration in regard to the relative system condition.